\documentclass[aps,prd,preprintnumbers,twocolumn,groupedaddress,superscriptaddress]{revtex4-1}
\usepackage{lipsum}
\usepackage{graphicx}
\usepackage{dcolumn}
\usepackage{bm}
\usepackage{amssymb}
\usepackage{amsmath}
\usepackage{xcolor}
\usepackage{amsmath}
\usepackage{amsfonts}
\usepackage{hyperref}
\hypersetup{colorlinks=true, linkcolor=blue, citecolor=blue, urlcolor=blue}
\usepackage{bm}
\usepackage{booktabs}

\begin{document}
\preprint{BARI-TH/780-25}

\title{Data-Driven Refinement of an Analytical Holographic Model for the QCD Phase Transition}
\author{Xun Chen}
\email{chenxun@usc.edu.cn}
\affiliation{School of Nuclear Science and Technology, University of South China, Hengyang 421001, China}
\affiliation{INFN --- Istituto Nazionale di Fisica Nucleare --- Sezione di Bari Via Orabona 4, 70125, Bari, Italy}

\author{Floriana Giannuzzi}
\email{floriana.giannuzzi@ba.infn.it}
\affiliation{INFN --- Istituto Nazionale di Fisica Nucleare --- Sezione di Bari Via Orabona 4, 70125, Bari, Italy}

\author{Stefano Nicotri}
\email{nicotri@infn.it}
\affiliation{INFN --- Istituto Nazionale di Fisica Nucleare --- Sezione di Bari Via Orabona 4, 70125, Bari, Italy}
\date{\today}

\begin{abstract}
Using (2+1)-flavor lattice QCD data, we refine the parameters of an analytical holographic model via gradient descent optimization to precisely locate the critical endpoint in the $T-\mu$ plane.
Specifically, we calibrate the model using input data for the speed of sound at $\mu_B = 0$, the second-order baryon number susceptibility $\chi^B_2$, and the baryon number density at $\mu_B/T = 1$.
With these parameters fixed, we calculate pressure and energy density versus temperature at small chemical potentials and compare the results with lattice QCD data using Taylor expansion techniques.
This comparison validates the robustness of our model upon extension to finite chemical potentials, as the results show broad consistency with lattice QCD data in this regime.
Finally, we employ the calibrated model to determine the coordinates of the critical endpoint in the $T-\mu_B$ plane, finding it located at $(\mu_B = 0.678 \, \rm{GeV}, T = 0.110 \, \rm{GeV})$.
\end{abstract}

\maketitle
\emph{Introduction}: The phase structure of strongly interacting matter described by QCD at finite temperature and baryon chemical potential is of fundamental importance in high-energy nuclear physics.
First-principles lattice QCD simulations have successfully established that, at vanishing chemical potential, the transition from hadronic matter to the quark-gluon plasma (QGP) is a smooth crossover for physical quark masses \cite{Aoki:2006we,Bazavov:2011nk,Borsanyi:2010bp}.
However, the sign problem prevents direct Monte Carlo calculations at finite baryon density, leaving the possible existence and location of a critical endpoint (CEP)--where the crossover turns into a first-order phase transition--an open question of significant theoretical and experimental interest \cite{Fukushima:2010bq, Stephanov:2004wx}.
A recent review of CEP can be found in Ref. \cite{Stephanov:2024xkn}.

Various approaches have been developed to explore the QCD phase diagram beyond the lattice-accessible region, including functional renormalization group methods (FRG) \cite{Fu:2023lcm, Osman:2024xkm,Roth:2024rbi,Murgana:2023pyx}, Dyson-Schwinger equations (DSE) \cite{Isserstedt:2019pgx,Wan:2024xeu,Lu:2023mkn}, and effective models like the Polyakov-loop-extended Nambu-Jona-Lasinio (PNJL) model \cite{Li:2018ygx,Liu:2023uxm,Bratovic:2012qs}.
Among these, holographic QCD models, based on the gauge/gravity duality, offer a compelling non-perturbative framework that remains applicable at strong coupling and finite density \cite{DeWolfe:2010he,Gursoy:2008bu,Ecker:2019xrw,Fu:2024nmw,Zhao:2022uxc,Grefa:2021qvt,Li:2024lrh,Zhu:2021nbl}.
In particular, Einstein-Maxwell-Dilaton (EMD) models have been used to successfully capture QCD thermodynamics, transport, and phase structure by incorporating dilaton and gauge field dynamics in a five-dimensional gravitational theory \cite{Critelli:2017oub,Yang:2017oer,Li:2020hau,Li:2025lmp,Li:2025ugv,Cao:2024jgt,Arefeva:2025uym,Arefeva:2025xtz,Chang:2024ksq,Hippert:2023bel,Chen:2018vty,Chen:2019rez,Chen:2019rez,Zhou:2020ssi,Chen:2021gop,Chen:2024ckb,Chen:2024epd,Chen:2024mmd,Chen:2024jet,Lin:2024mct,Zhu:2025gxo,Chen:2025ydu,Sun:2025uga,Zhang:2025wxi,Chen:2025kqb,Zhu:2025xiz,Chen:2025fpd,Shen:2025yrn,Jeong:2025omu}.

In this work, we develop a data-driven analytical EMD holographic model by calibrating its parameters against state-of-the-art (2+1)-flavor lattice QCD data at zero and finite baryon chemical potential.
Employing a gradient descent optimization scheme, we constrain the metric warp factor and gauge kinetic function to reproduce lattice observables such as the speed of sound, baryon number susceptibility, and baryon density.
In particular, at odds with previous studies, we fix parameters using also data at small values of chemical potential. We then extend the model to predict the QCD equation of state and locate the CEP in the $T$--$\mu$ plane.
Our results show quantitative agreement with lattice calculations and other holographic predictions, reinforcing the model's reliability in exploring the inaccessible regions of the QCD phase diagram.

\emph{Review of EMD framework}: We begin by reviewing the 5-dimensional Einstein-Maxwell-Dilaton system.
The action describes a gravitational field  $g_{\mu \nu}$, a Maxwell field $A_\mu$ and a dilaton field $\phi$.
Since we are only interested in the time component $A_0(z)=A_t(z)$, we set $A_i=0$ and make a gauge choice $A_z=0$. In the Einstein frame, the action is given by
\begin{equation}
    \begin{aligned}
        S_b = \frac{1}{16 \pi G_5} \int d^5 x \sqrt{-g} & \left[ R - \frac{f(\phi)}{4} F^2 \right. \\
        & \left. - \frac{1}{2} \partial_\mu \phi \partial^\mu \phi - V(\phi) \right],
    \end{aligned}
\end{equation}
where $f(\phi)$ is the gauge kinetic function describing the coupling between the dilaton $\phi$ and the Maxwell field $A_\mu$ (whose strength is $F^2$), $V\left(\phi\right)$ is the potential of the dilaton field, and $G_5$ is the Newton constant in five dimensions.
We use the metric ansatz
\begin{equation}
    d s^2=\frac{L^2 e^{2 A(z)}}{z^2}\left[-g(z) d t^2+\frac{d z^2}{g(z)}+d \vec{x}^2\right],
\end{equation}
where $z$ is the 5th-dimensional holographic coordinate and the radius $L$ of $\rm AdS_5$ space is set to unity.
Using the metric ansatz, the equations of motion (EOMs) and constraints for the background fields are:
\begin{equation}
    \begin{aligned}
        &\phi^{\prime \prime}+\phi^{\prime}\left(-\frac{3}{z}+\frac{g^{\prime}}{g}+3 A^{\prime}\right)-\frac{L^2 e^{2 A}}{z^2 g} \frac{\partial V}{\partial \phi}\\
        &+\frac{z^2 e^{-2 A} A_t^{\prime 2}}{2 L^2 g} \frac{\partial f}{\partial \phi}=0,
    \end{aligned}
\end{equation}
\begin{equation}
    A_t^{\prime \prime}+A_t^{\prime}\left(-\frac{1}{z}+\frac{1}{f}\frac{\mathrm{d}f}{\mathrm{d}z}+A^{\prime}\right)=0,
\end{equation}
\begin{equation}
    g^{\prime \prime}+g^{\prime}\left(-\frac{3}{z}+3 A^{\prime}\right)-\frac{e^{-2 A} A_t^{\prime 2} z^2 f}{L^2}=0,
\end{equation}
\begin{equation}
    \begin{aligned}
        A^{\prime \prime} & +\frac{g^{\prime \prime}}{6 g}+A^{\prime}\left(-\frac{6}{z}+\frac{3 g^{\prime}}{2 g}\right)-\frac{1}{z}\left(-\frac{4}{z}+\frac{3 g^{\prime}}{2 g}\right)\\
        &+3 A^{\prime 2} +\frac{L^2 e^{2 A} V}{3 z^2 g}=0,
    \end{aligned}
\end{equation}
\begin{equation}
    A^{\prime \prime}-A^{\prime}\left(-\frac{2}{z}+A^{\prime}\right)+\frac{\phi^{\prime 2}}{6}=0.
\end{equation}
Only four of the above five equations are independent.
The boundary conditions at the horizon are
\begin{equation}
    A_t\left(z_h\right)=g\left(z_h\right)=0.
\end{equation}
Near the asymptotic boundary $z \rightarrow 0$, we require the metric in the string frame to be asymptotic to $\rm AdS_5$.
The boundary conditions at $z = 0$ are:
\begin{equation}\label{eq:BC}
    A(0)=-\sqrt{\frac{1}{6}} \phi(0), \quad g(0)=1,
\end{equation}
and the boundary behavior of the Maxwell field is
\begin{equation}\label{eq:BCAt}
     A_t(z \rightarrow 0)=\mu+\rho^{\prime} z^2+\cdots,
\end{equation}
where $\mu$ denotes the baryon chemical potential, related to the quark chemical potential by $\mu = \mu_B = 3\mu_q$ and $\rho^{\prime}$ is a coefficient proportional to the baryon number density.
The density itself is calculated as \cite{Critelli:2017oub,Zhang:2022uin}
\begin{equation}
    \begin{aligned}
        \rho & =\left|\lim _{z \rightarrow 0} \frac{\partial \mathcal{L}}{\partial\left(\partial_z A_t\right)}\right| \\
        & =-\frac{1}{16\pi G_5} \lim _{z \rightarrow 0}\left[\frac{\mathrm{e}^{A(z)}}{z} f(\phi(z))  A_t'(z)\right],
    \end{aligned}
\end{equation}
where $\mathcal{L}$ is the Lagrangian density in the Einstein frame.
The equations of motion admit the following analytical expressions:
\begin{equation}
    \begin{aligned}
        \phi^{\prime}(z) & =\sqrt{-6\left(A^{\prime \prime}-A^{\prime\, 2}+\frac{2}{z} A^{\prime}\right)}, \\
        A_t(z) & =\sqrt{\frac{-1}{\int_0^{z_h} y^3 e^{-3 A} d y \int_{y_g}^y \frac{x}{e^{A} f} d x}} \int_z^{z_h} \frac{y}{e^{A} f} d y, \\
        g(z) & = 1 - \frac{\int_0^z y^3 e^{-3A} \left( \int_{y_g}^y \frac{x}{e^{A} f} dx \right) dy}{\int_0^{z_h} y^3 e^{-3A} \left( \int_{y_g}^y \frac{x}{e^{A} f} dx \right) dy}, \\
        V(z) & =\frac{-3 z^2 g \, e^{-2 A}}{L^2}\bigg[A^{\prime \prime} + 3 A^{\prime \, 2}+ \frac{g^{\prime \prime}}{6 g} + \\
        & + \left(\frac{3 g^{\prime}}{2 g}-\frac{6}{z}\right) A^{\prime} - \frac{1}{z}\left(\frac{3 g^{\prime}}{2 g}-\frac{4}{z}\right)\bigg].
    \end{aligned}
\end{equation}
The integration constant $y_g$ is related to the chemical potential $\mu$ by the boundary condition in \eqref{eq:BCAt}, yielding:
\begin{equation}\label{eq:mu}
    \mu=\sqrt{\frac{-1}{\int_0^{z_h} y^3 e^{-3 A(y)} (\int_{y_g}^y \frac{x}{e^{A(x)} f(x)} d x) d y }} \int_0^{z_h} \frac{y}{e^{A(y)} f(y)}d y.
\end{equation}
We adopt a general ansatz for the gauge kinetic function:
\begin{equation}\label{fff}
    f(z)=\frac{k\,  e^{c z^2+ h z^4 + n z^6 -A(z)}}{-2c- 4 h z^2-6 n z^4}.
\end{equation}
In this case, Eq. \eqref{eq:mu} simplifies and the relation between the integration constant $y_g$ and the chemical potential $\mu$ becomes:
\begin{equation}
    e^{-c y_g^2-h y_g^4-n y_g^6} = \frac{I_2(0,z_h)}{I_1(0,z_h)} + \frac{(e^{-c z_h^2-h z_h^4-n z_h^6}-1)^2}{k\,\mu^2 I_1(0,z_h)},
\end{equation}
where
\begin{equation}
    \begin{aligned}
        I_1(a,b) &= \int_a^b y^3 e^{-3A(y)}\,dy \\
        I_2(a,b) &= \int_a^b y^3 e^{-3A(y)-c y^2-h y^4-n y^6}\,dy.
    \end{aligned}
\end{equation}
Consequently, the solutions can be written in the following simplified forms:
\begin{equation}
    \begin{aligned}
        g(z)&=1-\frac{1}{I_1(0,z_h)}\!\left[I_1(0,z) -\frac{\mu^2 k}{\Delta^{2}}\,\det\mathcal G\right],\\
        \phi^{\prime}(z) & =\sqrt{6\left(A^{\prime 2}-A^{\prime \prime}-2 A^{\prime} / z\right)}, \\
        A_t(z)&=\mu\ \frac{e^{-c z^2-h z^4 - n z^6}-e^{-c z_h^2-h z_h^4 - n z_h^6}}{1-e^{-c z_h^2-h z_h^4 - n z_h^6}}, \\
        V(z) & =\frac{-3 z^2 g \, e^{-2 A}}{L^2}\bigg[A^{\prime \prime} + 3 A^{\prime \, 2}+ \frac{g^{\prime \prime}}{6 g} + \\
        & + \left(\frac{3 g^{\prime}}{2 g}-\frac{6}{z}\right) A^{\prime} - \frac{1}{z}\left(\frac{3 g^{\prime}}{2 g}-\frac{4}{z}\right)\bigg],
    \end{aligned}
\end{equation}
with
\begin{equation}
    \Delta = e^{-c z_h^2-h z_h^4 - n z_h^6} - 1 ,
\end{equation}
and
\begin{equation}
    \det\mathcal G=
    \begin{vmatrix}
        I_1(0,z_h) & I_2(0,z_h)\\[2mm]
        I_1(z_h,z) & I_2(z_h,z)
    \end{vmatrix}.
\end{equation}
The Hawking temperature and entropy of this black hole solution are given by
\begin{equation}
T = \frac{z_h^3 e^{-3A(z_h)}}{4\pi I_1(0,z_h)} \left[ 1 - \frac{\mu^2 k}{\Delta^2} \Bigl( (\Delta + 1) I_1(0,z_h) - I_2(0,z_h) \Bigr) \right]
\end{equation}
and
\begin{equation}\label{SSS}
    s=\frac{e^{3 A\left(z_h\right)}}{4 G_5 z_h^3}.
\end{equation}
To obtain an analytical solution, we assume the form
\begin{equation}\label{AAA}
    A(z)= d*\ln(a z^2 + 1) + d*\ln(b z^4 + 1).
\end{equation}
The corresponding metric function in the string frame is $A_s(z)=A(z)+\sqrt{\frac{1}{6}} \phi(z)$.
The first four parameters ($a,b,d,G_5$) will be fixed by matching the equation of state obtained in the holographic model with lattice QCD data using a machine-learning approach, as described in the next section.
The other parameters ($c$, $k$, $h$ and $n$) will be constrained by the baryon number susceptibility and baryon number density.
Once the entropy and the density $\rho$ are determined, the free energy and pressure are calculated integrating the thermodynamic relation:
\begin{equation}
    \begin{aligned}
        dF&= -dp = - s\, d T- \rho\, d \mu
    \end{aligned}
\end{equation}
as in \cite{Chen:2024mmd,Critelli:2017oub}.
We have normalized the free energy to vanish at $(T,\mu)=(0,0)$.
Then, the energy density of the system can be derived as
\begin{equation}
\epsilon=-p+s\, T+\mu\, \rho.
\end{equation}
The specific heat at constant volume is \cite{DeWolfe:2010he}
\begin{equation}
    C_V=T \frac{\partial s}{\partial T}.
\end{equation}
At non-zero chemical potential, the squared speed of sound can be calculated by \cite{Li:2020hau,Yang:2017oer,Gursoy:2017wzz}
\begin{equation}
    C_s^2=\frac{s}{T\left(\frac{\partial s}{\partial T}\right)_\mu+\mu\left(\frac{\partial \rho}{\partial T}\right)_\mu},
\end{equation}
and the second-order baryon number susceptibility is defined as
\begin{equation}
    \chi_2^B=\frac{1}{T^2} \frac{\partial \rho}{\partial \mu}.
\end{equation}

\emph{Thermodynamics and phase diagram}:
The model is calibrated through an iterative process against the lattice results.
Specifically, we use a gradient descent optimizer to automatically adjust the model parameters, thereby minimizing the difference between the holographic outcomes and lattice calculations.
Further details can be found in Refs. \cite{Chen:2024ckb,Chen:2024mmd}. We calibrated our model using lattice QCD data for the following observables: the speed of sound at $\mu_B = 0$, the second-order baryon number susceptibility $\chi^B_2$, and the baryon number density at $\mu_B/T = 1$  \cite{HotQCD:2014kol,HotQCD:2012fhj,Bazavov:2017dus}.
The values of the parameters obtained from this calibration are shown in Table~\ref{tab:parameters}.
\begin{table}[htbp]
    \caption{Optimized model parameters from lattice QCD calibration.}
    \label{tab:parameters}
    \begin{ruledtabular}
    \begin{tabular}{cccc}
        Parameter & Value & Parameter & Value \\
        \hline
        $a$ & 0.2219 GeV$^{2}$  & $k$ & 0.02243 GeV$^{2}$ \\
        $b$ & 0.0330 GeV$^{4}$ & $n$ & -0.0010 GeV$^{6}$ \\
        $c$ & -0.0358 GeV$^{2}$ & $h$ & 0.0039 GeV$^{4}$ \\
        $d$ & -0.1212 & $G_5$ & 0.4113  \\
    \end{tabular}
    \end{ruledtabular}
\end{table}

\begin{figure*}
    \centering
    \includegraphics[width=1\textwidth]{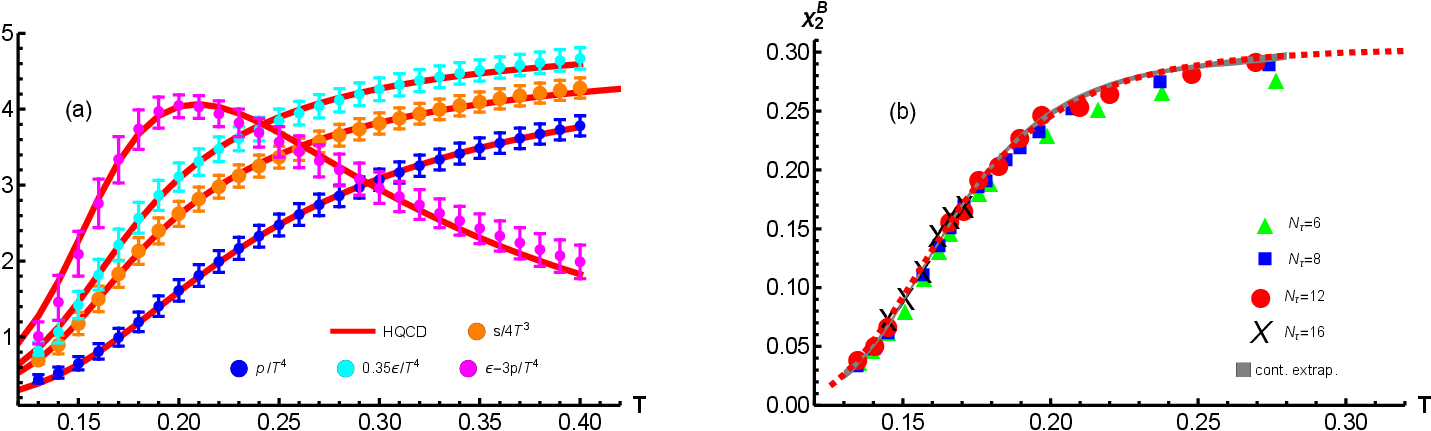}
    \caption{\label{eoschi} Comparison of (2+1)-flavor lattice QCD \cite{HotQCD:2014kol,HotQCD:2012fhj,Bazavov:2017dus} and holographic model predictions for: (a) thermodynamic quantities (entropy $s$, energy $\epsilon$, pressure $p$, trace anomaly $\epsilon - 3p$) versus temperature at vanishing chemical potential; (b) second-order baryon number susceptibility versus temperature at vanishing chemical potential.}
\end{figure*}

The thermodynamic behavior of the system at vanishing baryon chemical potential is summarized in Fig.~\ref{eoschi}.
Panel (a) displays the temperature dependence of key thermodynamic quantities: entropy density $s$, energy density $\epsilon$,  pressure $p$, and the trace anomaly $\epsilon - 3p$, all normalized by a proper power of $T$, according to their scaling at large temperature.
The predictions of the holographic model (solid red curves) show remarkable agreement with lattice QCD data across the entire temperature range.
Particularly noteworthy is the model's ability to capture the rapid rise of these quantities during the crossover transition from the confined to the deconfined phase.
The trace anomaly, which serves as a sensitive indicator of the transition region, exhibits a characteristic peak that is well reproduced by the holographic approach.
This agreement demonstrates that our model correctly captures the non-perturbative dynamics near the transition temperature $T_c$.
At high temperatures ($T \gtrsim 2T_c$), all quantities approach their respective Stefan-Boltzmann limits \cite{Borsanyi:2010cj}, as expected  for a weakly interacting quark-gluon plasma.
The holographic model smoothly interpolates between the strongly coupled regime near $T_c$ and the weakly coupled limit at high temperatures, providing a unified description of the equation of state.
Panel (b) of Fig.~\ref{eoschi} shows the second-order baryon number susceptibility $\chi_2^B$, which measures the response of the baryon density to changes in the chemical potential and serves as a probe of the relevant degrees  of freedom in the system.
The comparison includes lattice QCD data with different temporal extents $N_\tau$ (various symbols) as well as continuum extrapolated results (gray squares).

\begin{figure*}
    \centering
    \includegraphics[width=1\textwidth]{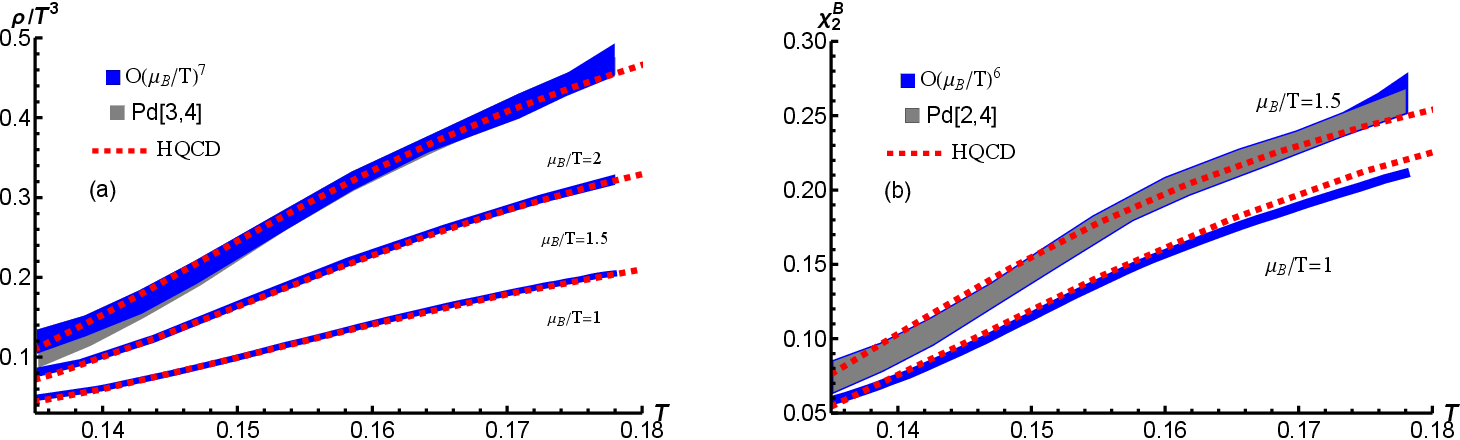}
    \caption{\label{rhochi} Comparison of (2+1)-flavor lattice QCD \cite{Bollweg:2022rps} and holographic model predictions for: (a) baryon number density versus temperature at finite chemical potential; (b) second-order baryon number susceptibility versus temperature at finite chemical potential.}
\end{figure*}

Fig. ~\ref{rhochi} (a) displays the temperature dependence of the baryon number density  $\rho/T^3$ at finite chemical potential $\mu_B$, with results shown for several representative values of $\mu_B/T$.
Fig. ~\ref{rhochi} (b) displays the temperature dependence of second-order baryon number susceptibility  $\chi_2^B$ for two representative values of $\mu_B/T$.
The predictions of the holographic model (HQCD, red dashed curves) are compared with lattice QCD results obtained from $[n,4]$ Pad\'{e}
 approximants (gray bands) and conventional Taylor expansions truncated at $O((\mu_B/T)^7)$ (blue bands) \cite{Bollweg:2022rps}.
The HQCD model successfully captures the characteristic increase in baryon density with increasing chemical potential across the entire temperature range.

\begin{figure*}
    \centering
    \includegraphics[width=1\textwidth]{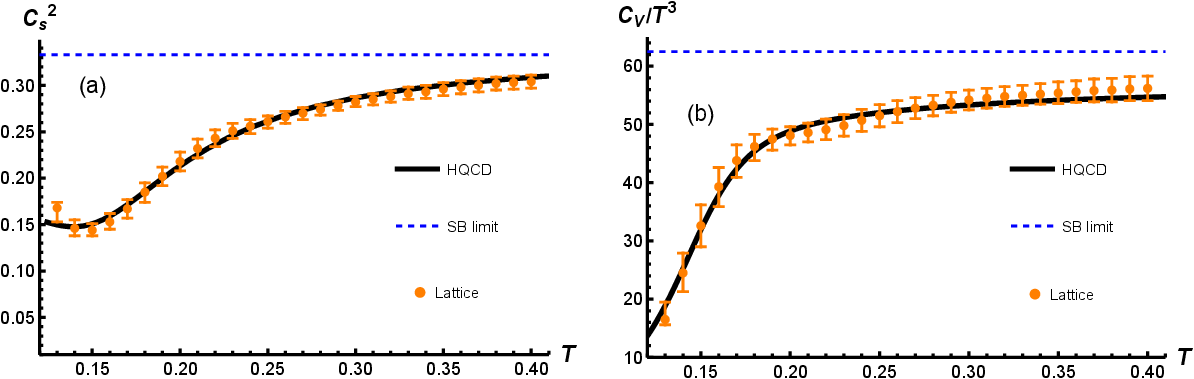}
    \caption{\label{cscv} Comparison between (2+1)-flavor lattice QCD \cite{HotQCD:2014kol} and holographic model predictions at zero chemical potential for: (a) temperature dependence of the squared speed of sound; (b) temperature dependence of the specific heat.}
\end{figure*}

Fig.~\ref{cscv} (a) shows the temperature dependence of the squared speed of sound $C_s^2$ at $\mu_B=0$, which provides crucial information about the conformal behavior and stiffness of the equation of state.
The holographic model (solid black curve) exhibits a smooth, monotonic increase from $C_s^2 \approx 0.15$ at $T = 150$ MeV to approximately $0.30$ at
$T = 400$ MeV, indicating a gradual approach toward the conformal limit of $C_s^2 = 1/3$ \cite{Borsanyi:2010cj}.
The excellent agreement between the HQCD predictions and lattice QCD data (orange points) demonstrates that the model correctly captures the restoration of conformal symmetry at high temperatures.
The smooth progression without any discontinuities is consistent with the crossover nature of the QCD transition at physical quark masses.
Particularly noteworthy is the model's ability to reproduce the low-temperature behavior where $C_s^2$ exhibits a minimum near the transition region, reflecting the softest point  in the equation of state.
The specific heat $C_V/T^3$, shown in Fig.~\ref{cscv} (b), serves as a sensitive probe of thermodynamic fluctuations and degrees of freedom in the system. The comparison with lattice QCD data reveals generally good agreement, particularly in the high-temperature regime where the Stefan-Boltzmann limit $C_V/T^3 \approx 62.5$ \cite{Borsanyi:2010cj} is gradually approached.

\begin{figure*}
    \centering
    \includegraphics[width=1\textwidth]{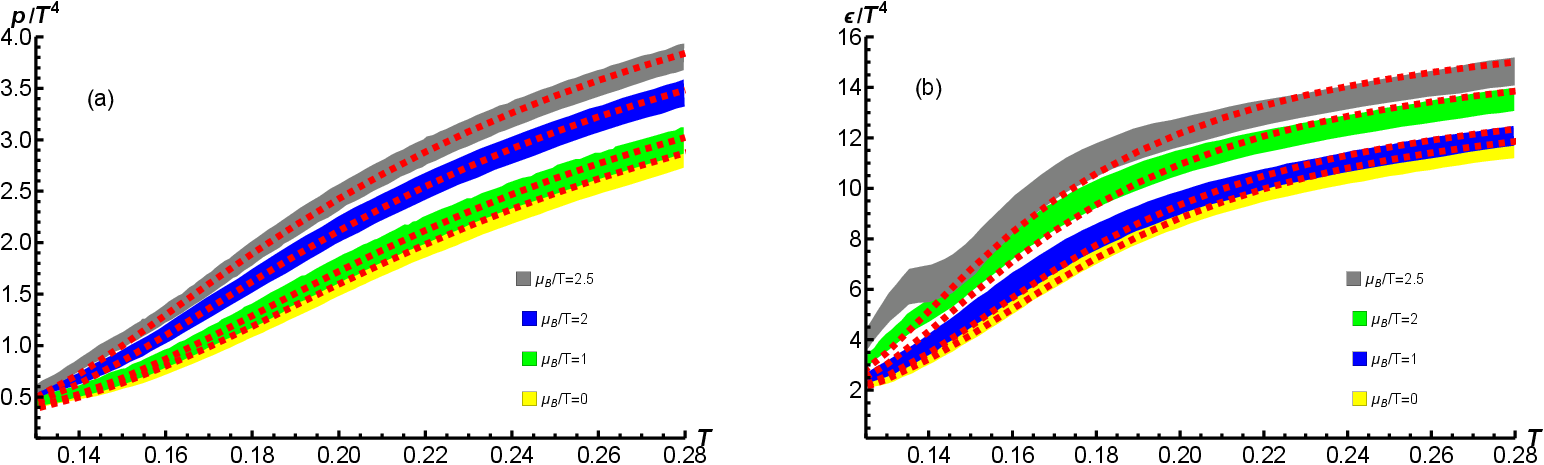}
    \caption{\label{PEmu} (a) Comparison of pressure between our holographic model and lattice QCD data at various $\mu/T$ values \cite{Bazavov:2017dus}; (b) Comparison of energy density $\epsilon$ between model predictions and lattice results at various $\mu/T$ values \cite{Bazavov:2017dus}.}
\end{figure*}

Fig.~\ref{PEmu} (a) displays the temperature dependence of the normalized pressure $p/T^4$ at finite chemical potential.
The holographic model shows good agreement with lattice QCD data across all values of $\mu_B/T$ investigated.
The pressure exhibits a characteristic rapid increase in the transition region.
A key observation is the enhancement of $p/T^4$ with increasing chemical potential $\mu_B/T$, particularly noticeable in the intermediate temperature range.
This  enhancement reflects the additional contribution from finite baryon density to the equation of state.
The holographic model successfully captures this dependence, demonstrating its capability to describe the thermodynamic properties of dense QCD matter.
The smooth behavior of the pressure without any discontinuities is consistent with the crossover nature of the QCD transition at physical quark masses and at such small values of chemical potential.

The normalized energy density $\epsilon/T^4$, shown in Fig.~\ref{PEmu} (b), provides complementary information about the equation of state.
As in the case of the pressure, the holographic model predictions agree with lattice QCD results (some discrepancies are found at low $T$).
The holographic model correctly reproduces the enhancement of $\epsilon/T^4$ with increasing $\mu_B/T$.

\begin{figure*}
    \centering
    \includegraphics[width=1\textwidth]{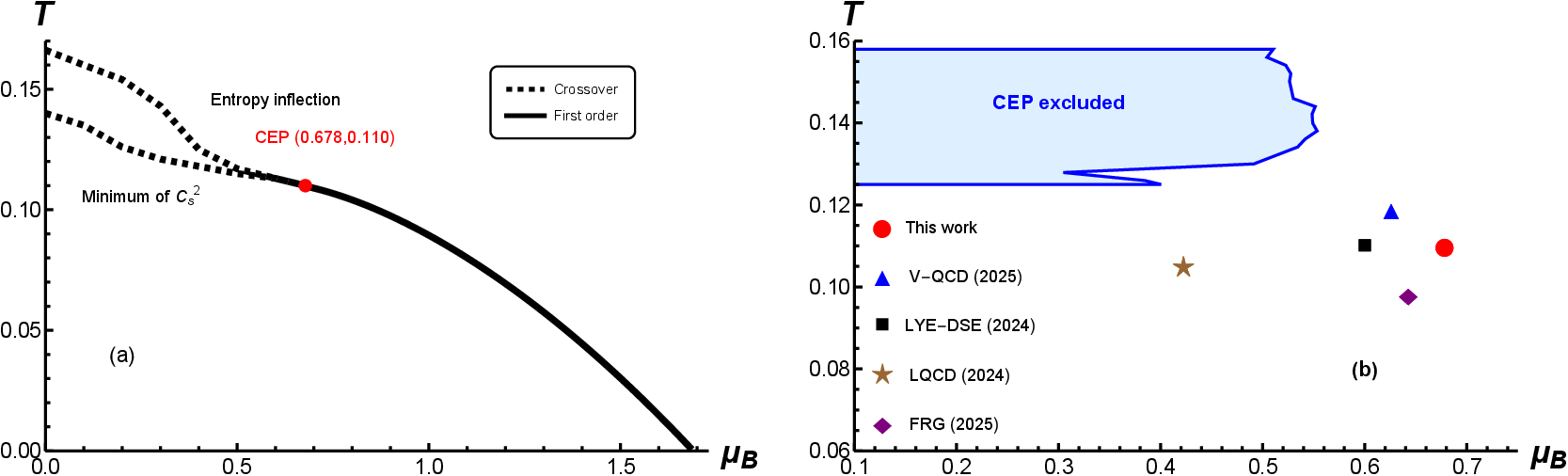}
    \caption{\label{phasediagram}  (a) The QCD phase diagram in the $T$-$\mu$ plane; (b) Comparison of our model predictions with those of other theoretical models, including: the CEP-excluded region from lattice QCD \cite{Borsanyi:2025dyp}; the CEP position from Lattice QCD ($\mu_{B}^{\mathrm{CEP}} = 422_{-35}^{+80}~\mathrm{MeV}$, $T^{\mathrm{CEP}} = 105_{-18}^{+8}~\mathrm{MeV}$) \cite{Clarke:2024ugt}; from V-QCD ($\mu_{B}^{\mathrm{CEP}} = 626_{-179}^{+90}~\mathrm{MeV}$, $T^{\mathrm{CEP}} = 105_{-6}^{+14}~\mathrm{MeV}$) \cite{Ecker:2025vnb}; from LYE-DSE \cite{Wan:2024xeu}; and from FRG \cite{Fu:2023lcm}.}
\end{figure*}

\begin{figure*}
    \centering
    \includegraphics[width=.4\textwidth]{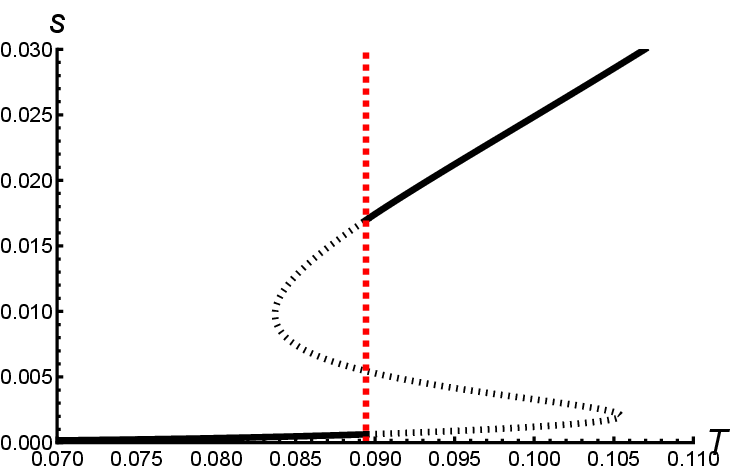}
    \caption{Entropy density (in GeV$^3$) at $\mu=1$ GeV as a function of temperature (in GeV). At $T\sim 89.4$ MeV a first order phase transition occurs.}
    \label{fig:entropyjump}
\end{figure*}

Fig.~\ref{phasediagram} (a) presents the QCD phase diagram in the temperature-chemical potential ($T$-$\mu$) plane as predicted by this holographic model.
The diagram clearly shows the characteristic features of the QCD phase structure: a crossover transition at small chemical potentials  that evolves into a first-order phase transition at higher densities.
We have chosen two definitions of the transition line in the crossover region, producing the two dashed lines shown in the figure.
The upper line is obtained finding the inflection points of the $s/T^3$ versus $T$ curves at fixed values of the chemical potential.
The other line is obtained locating the minimum of the squared speed of sound as a function of temperature at fixed chemical potential.
At values of the chemical potential higher than a critical one, the relation between temperature and $z_h$ is no more a bijection and a first-order phase transition occurs \cite{Colangelo:2013ila}.
In this case, a jump in some thermodynamic functions, as e.g. entropy and baryon number density, is found at the critical temperature, as shown in Fig. \ref{fig:entropyjump}.
The first-order transition line has been computed finding, for each value of the chemical potential, the temperature at which the free energy is the same in the two phases \cite{DeWolfe:2010he}.
Analogously, we have found the same result with a different procedure, determining, at fixed chemical potential, the temperature $T_f$ at which the closed regions bounded by the entropy density on either side of the $T=T_f$ line have equal area (for more details see \cite{DeWolfe:2010he}).
The critical endpoint (CEP), marked by the red circle at $(T_{\text{CEP}}, \mu_{\text{CEP}}) = (0.110, 0.678)$ GeV, represents the termination point of the first-order transition line.
At this point, the phase transition becomes second order, with associated critical phenomena. The location of the CEP in our model falls within the theoretically expected range and is consistent with constraints from first-principles calculations.
The smooth connection between the crossover and first-order transition regions demonstrates
the model's ability to describe the entire phase structure consistently.
As expected in chiral and deconfinement transitions, the transition temperature decreases as the chemical potential increases.
Fig.~\ref{phasediagram} (b) provides a comparison of our model predictions with results from other theoretical approaches.
This convergence suggests an emerging consensus within the holographic frameworks regarding the location of the critical endpoint.
Especially, the CEP position predicted by the holographic V-QCD model \cite{Ecker:2025vnb}, whose parameters have been fixed from neutron star observations, is consistent with the result found in this model, constrained from  lattice QCD.
This agreement from two independent lines of evidence reinforces the validity of our model.

\emph{Conclusion}: In this work, we have established a refined analytical holographic model for QCD thermodynamics by systematically calibrating it against state-of-the-art (2+1)-flavor lattice QCD data with machine learning.
The model successfully describes the equation of state across a wide range of temperatures and baryon chemical potentials, demonstrating remarkable agreement with first-principles calculations.
The improved agreement with lattice computations at zero and small chemical potential with respect to previous studies has been achieved by introducing two more parameters is the gauge kinetic function $f(\phi)$, and by calibrating the model using input data for the speed of sound at zero chemical potential, and also data at finite but small chemical potential concerning second-order baryon number susceptibility and baryon number density.

The core achievement of this study is the precise determination of the QCD phase diagram, as summarized in Fig.~\ref{phasediagram} (a).
Our model predicts a smooth crossover transition at low chemical potentials that evolves into a first-order phase transition at higher densities, culminating in a critical endpoint (CEP) located at $(\mu_{\text{CEP}}, T_{\text{CEP}}) = (0.678\ \text{GeV}, 0.110\ \text{GeV})$. This prediction is positioned within the broader landscape of theoretical approaches in Fig.~\ref{phasediagram} (b), where it shows strong consistency with some recent results from other holographic QCD studies.

Beyond the phase structure, the model exhibits robust performance in reproducing fundamental and derived thermodynamic quantities.
It accurately captures the behavior of the entropy density, pressure, energy density, and trace anomaly at $\mu_B=0$, as well as the second-order baryon number susceptibility $\chi_2^B$.
Notably, at small chemical potential, our holographic predictions for the baryon number density show agreement with Pad\'{e}-resummed lattice data and truncated Taylor expansions.

The comprehensive agreement between our holographic framework and lattice QCD data validates the use of such bottom-up models as a powerful and consistent tool for extrapolating into regions of the QCD phase diagram that are inaccessible to standard lattice simulations due to the sign problem.
The successful description of the equation of state, along with the precise location of the CEP, provides valuable theoretical guidance for ongoing experimental searches in heavy-ion collision programs and contributes to a more profound understanding of dense matter in neutron stars.

\section*{Acknowledgments}
This work is supported  by the National Natural Science Foundation of China (NSFC) Grant Nos: 12405154, and the European Union --- Next Generation EU through the research grant number P2022Z4P4B ``SOPHYA --- Sustainable Optimised PHYsics Algorithms: fundamental physics to build an advanced society'' under the program PRIN 2022 PNRR of the Italian Ministero dell'Universit\`a e Ricerca (MUR).
We thank P.~Colangelo and F.~De~Fazio for fruitful discussions.
\bibliography{ref}
\end{document}